\documentclass[prb,aps,twocolumn,superscriptaddress,longbibliography]{revtex4-2}
\usepackage{graphicx,color}
\usepackage{amsthm}
\usepackage{amsfonts}
\usepackage{algorithmic}
\usepackage{enumerate}
\usepackage{latexsym}
\usepackage{amsmath}
\usepackage{amssymb}
\usepackage{multirow}
\usepackage{array}
\usepackage{diagbox}
\usepackage[normalem]{ulem}
\usepackage{subfigure}
\usepackage[colorlinks=true,citecolor=blue,linkcolor=blue]{hyperref}

\emergencystretch=\maxdimen
\makeatletter
\DeclareFontFamily{OMX}{MnSymbolE}{}
\DeclareSymbolFont{MnLargeSymbols}{OMX}{MnSymbolE}{m}{n}
\SetSymbolFont{MnLargeSymbols}{bold}{OMX}{MnSymbolE}{b}{n}
\DeclareFontShape{OMX}{MnSymbolE}{m}{n}{
    <-6>  MnSymbolE5
   <6-7>  MnSymbolE6
   <7-8>  MnSymbolE7
   <8-9>  MnSymbolE8
   <9-10> MnSymbolE9
  <10-12> MnSymbolE10
  <12->   MnSymbolE12
}{}
\DeclareFontShape{OMX}{MnSymbolE}{b}{n}{
    <-6>  MnSymbolE-Bold5
   <6-7>  MnSymbolE-Bold6
   <7-8>  MnSymbolE-Bold7
   <8-9>  MnSymbolE-Bold8
   <9-10> MnSymbolE-Bold9
  <10-12> MnSymbolE-Bold10
  <12->   MnSymbolE-Bold12
}{}

\let\llangle\@undefined
\let\rrangle\@undefined
\DeclareMathDelimiter{\llangle}{\mathopen}%
                     {MnLargeSymbols}{'164}{MnLargeSymbols}{'164}
\DeclareMathDelimiter{\rrangle}{\mathclose}%
                     {MnLargeSymbols}{'171}{MnLargeSymbols}{'171}
\makeatother

\NewDocumentCommand{\dgal}{sO{}m}{%
  \IfBooleanTF{#1}
    {\dgalext{#3}}
    {\dgalx[#2]{#3}}%
}
\NewDocumentCommand{\dgalext}{m}{%
  \sbox0{%
    \mathsurround=0pt 
    $\left\{\vphantom{#1}\right.\kern-\nulldelimiterspace$%
  }%
  \sbox2{\{}%
  \ifdim\ht0=\ht2
    \{\kern-.625\wd2 \{#1\}\kern-.625\wd2 \}%
  \else
    \left\{\kern-.7\wd0\left\{#1\right\}\kern-.7\wd0\right\}%
  \fi
}

\begin{document}
\title{Dual role of stripe phase on superconducting correlation in a bilayer square lattice}
\author{Ting Guo}
\affiliation{School of Physics and Astronomy, Beijing Normal University, Beijing 100875, China\\}
\author{Lufeng Zhang}
\affiliation{School of Science, Beijing University of Posts and Telecommunications, Beijing 100876, China\\}
\affiliation{Key Laboratory of Multiscale Spin Physics (Ministry of Education), Beijing Normal University, Beijing 100875, China}
\author{Tianxing Ma}
\email{txma@bnu.edu.cn}
\affiliation{School of Physics and Astronomy, Beijing Normal University, Beijing 100875, China\\}
\affiliation{Key Laboratory of Multiscale Spin Physics (Ministry of Education), Beijing Normal University, Beijing 100875, China}
\author{Hai-Qing Lin}
\affiliation{ School of Physics and Institute for Advance Study in Physics, Zhejiang University, Hangzhou 310058, Zhejiang, China \\}

\begin{abstract}
While the stripe phase has been observed not only in monolayer cuprates but also in bilayer cuprates, research on its behavior in bilayer cuprates has been limited. Using constrained path quantum Monte Carlo, we explore the effect of stripes on the bilayer square lattice. We find the system exhibits short-range antiferromagnetism, which is enhanced by stripes and is strongest when the electron density of the interstriped rows reaches half-filling. The hole doping concentration plays a crucial role in the interaction between stripes and superconductivity. The $d$-wave pairing is enhanced by stripe potential $V_0$ at the hole doping $\delta_h=1/4$, whereas it is suppressed by stripe potential $V_0$ at the hole doping $\delta_h=1/8$. We elucidate this phenomenon through an analysis of the magnetism of the interstriped rows. Furthermore, the effective $d$-wave pairing is stronger in the bilayer model compared to the monolayer model when stripes are introduced on the square lattice. Overall, our unbiased numerical simulations provide a further understanding of the crossed bilayer square lattice model.
\end{abstract}
\maketitle

\section{Introduction}
High-temperature superconductors are a remarkable achievement of modern physics, and revealing the microscopic mechanisms underlying high-temperature superconductivity has been a long-standing challenge and a focal point of physics research \cite{Bednorz:1986tc,zhou2021high}. To achieve a comprehensive understanding of the microscopic mechanisms of high-temperature superconductors, it is crucial to grasp the prevalent symmetry breaking, which is typically manifested as charge density waves(CDWs) \cite{Tranquada1995EvidenceFS,PhysRevLett.129.027002,doi:10.1126/science.1223532,Kawasaki2024,Tabis2014,Aishwarya2023}. However, the relationship between CDW and unconventional superconductivity is still controversial. For instance, the anomalous suppression of superconductivity observed at a hole doping concentration of close to 1/8 has been shown to correlate with charge and spin stripe order in the LaSrCuO system \cite{Tranquada1995EvidenceFS,RevModPhys.75.1201}. In YBCO materials, application of a strong magnetic field simultaneously suppresses superconductivity and enhances CDW order, thereby revealing a clear competitive interaction between CDW order and superconductivity \cite{Chang2016,Choi2020}. In Kagome superconductor $\text{CsV}_3\text{Sb}_5$ under moderate pressure conditions, CDW order is monotonically suppressed by pressure, while the superconductivity shows a two-dome-like behavior, suggesting that the interplay between CDW order and superconductivity is more intricate than simple competition \cite{Zheng2022}. Furthermore, numerous theoretical frameworks and simulations have suggested that CDW play a positive role in enhancing pairing correlations \cite{PhysRevB.86.184506,doi:10.1073/pnas.2109406119,doi:10.1126/science.adh7691,PhysRevLett.75.4650,PhysRevB.96.064527,PhysRevLett.104.247001,PhysRevB.105.155154}. For example, dynamic cluster quantum Monte Carlo simulations indicate that imposing stripes can lead to a significant increase in $d$-wave pairing when the modulation scale is sufficiently large \cite{PhysRevLett.104.247001}.

CDW in high-temperature superconductors exhibit a variety of forms, including stripe phase \cite{Tranquada1995EvidenceFS,PhysRevLett.78.338}, checkerboard pattern \cite{Hanaguri2004,PhysRevResearch.2.042042} and chiral CDW \cite{Song2022,Jiang2021}, among others. Additionally, in certain circumstances, the observed CDW may be a composite or hybrid form of these fundamental patterns \cite{Ning2024,Wu2024}. The stripe phase, as a basic form of CDW, features a relatively simple geometric structure and is found in wide range of diverse materials \cite{Jin2021,doi:10.1126/science.aak9546,doi:10.1126/science.1258399,Zhao2019}, making it an ideal model for studying CDW phenomena. The relevant research on the stripe phase has mainly focused on specific two-dimensional planes \cite{PhysRevB.103.155110,PhysRevLett.117.046603,PhysRevB.103.155110,PhysRevB.105.035111,PhysRevB.109.024505,PhysRevB.106.235140,PhysRevLett.88.117001,PhysRevB.76.140505}, and in particular on the monolayer square lattice \cite{PhysRevLett.113.046402,PhysRevX.11.031007,PhysRevB.105.115116,PhysRevB.109.045101,Chen2024}. For instance, by utilizing advanced computational approaches that integrate the strengths and weaknesses of multiple numerical tools, Zheng et al. uncovered the stripe period in the underdoped ground state of the monolayer square lattice under medium-to-strong coupling conditions and demonstrated the coupling of stripe fluctuations with pairing order \cite{doi:10.1126/science.aam7127}. Nevertheless, studies on the stripe phase in higher-dimensional systems remain relatively scarce, and further advances are awaited\cite{PhysRevB.108.165131}.

Different types of cuprate superconductors have varying numbers of CuO$_2$ planes per unit cell. Notably, the stripe phase has been observed not only in monolayer compounds but also in multi-layer compounds \cite{doi:10.1126/science.1258399,Zhao2019,PhysRevLett.99.127003}. Furthermore, compounds with double or multiple CuO$_2$ layers typically exhibit higher superconducting transition temperatures($T_c$) than monolayer compounds \cite{osti_6456074}. In fact, the cuprate superconductor with the currently known highest $T_c$, Hg-Ba-Ca-Cu-O(1223), features a trilayer copper-oxygen structure \cite{PhysRevB.50.4260}. This intriguing phenomenon raises several important questions: Does the stripe phase cause the difference between monolayer and multi-layer cuprates? What is the specific impact of the stripe phase within these multi-layer cuprates? Therefore, we employ a bilayer square lattice to investigate the effect of the stripe phase in fundamental multi-layer cuprates —bilayer cuprates.

This study uses the Constrained Path quantum Monte Carlo(CPMC) method with the extended Hubbard model to study the stripe phase in a bilayer square lattice. The main conclusions are as follows: Applying charge modulation enhances the antiferromagnetic properties of the system and generates a ${\pi}$-phase shift. The hole doping concentration plays a crucial role in the interaction between stripes and superconductivity. The $d$-wave pairing is enhanced by stripe potential $V_0$ at the hole doping $\delta_h=1/4$ but is suppressed by stripe potential $V_0$ at the hole doping $\delta_h=1/8$. In regions without stripes or with weak stripe potential $V_0$, the $d$-wave pairing in a monolayer lattice is stronger than in a bilayer lattice. However, in regions with strong stripe potential $V_0$, the $d$-wave pairing in a bilayer lattice surpasses that in a monolayer lattice. The existence of additional nearest neighbor attractive interactions causes an enhancement of the $d$-wave pairing.

\section{Model and methods}
\begin{figure}[tbp]
\includegraphics[scale=0.35]{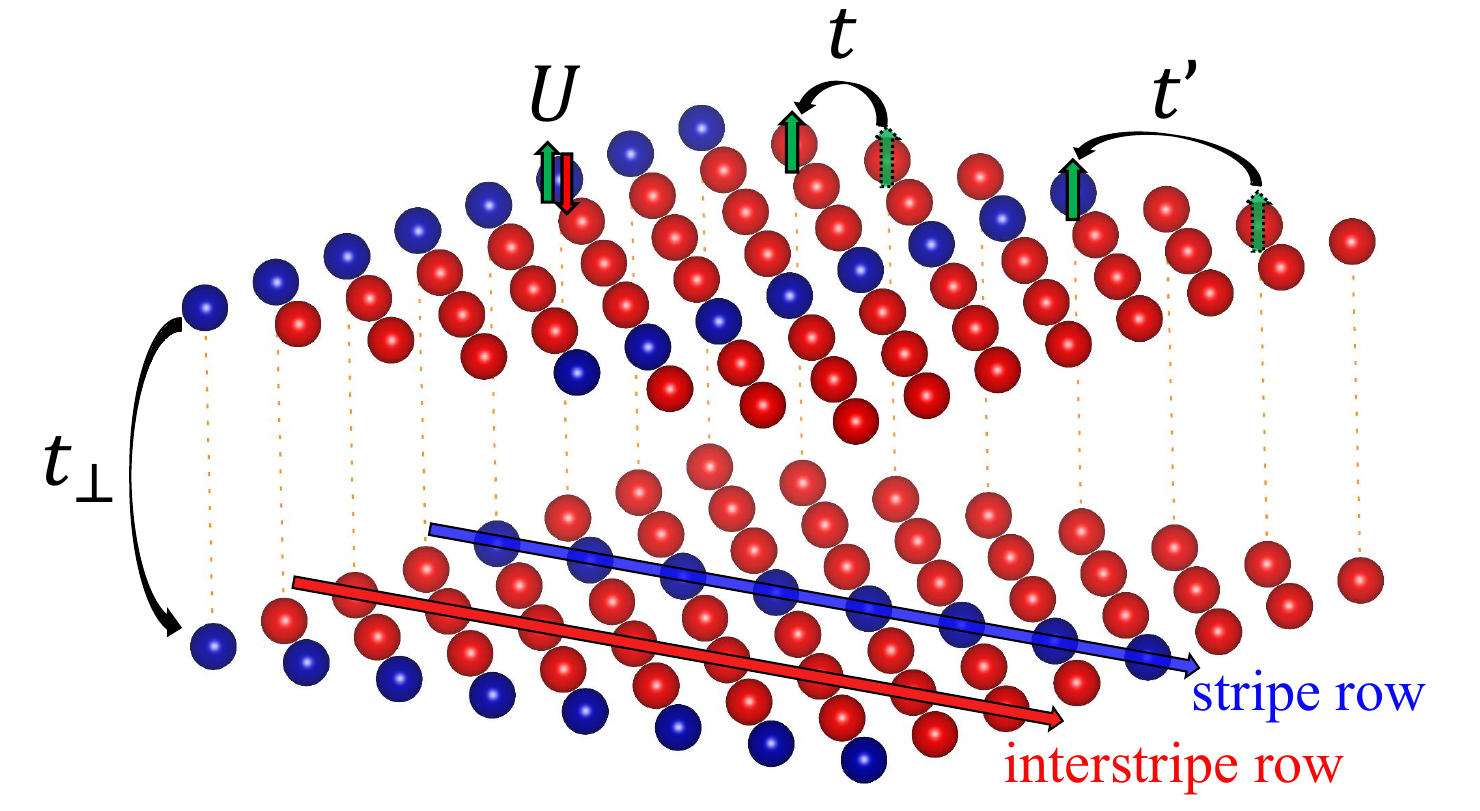}
\caption{ Illustration of the crossed bilayer square lattice with period P = 4. The blue filled spheres represent the sites with stripe potential $V_0$, whereas the red filled spheres represent the sites are without stripe potential $V_0$. The blue filled spheres label the striped rows. $t$, $t'$ and $t_{\perp}$ denote hopping parameters. $U$ is the onsite Coulomb repulsion interaction.
}
\label{Fig1}
\end{figure}

We study the effect of charge stripes on a bilayer square lattice. There are different types of the striped bilayer square lattice due to the existence of charge stripes. In this article, we mainly consider the crossed bilayer square lattice model, which is in agreement with numerical simulations and experiments \cite{doi:10.1126/science.1258399,PhysRevB.71.220504,https://doi.org/10.1002/adma.202307515}. In the crossed bilayer square lattice model, the striped rows of the first layer are perpendicular to the striped rows of the second layer and the $2\times8\times8$ lattice is shown in Fig. \ref{Fig1}. The blue filled spheres represent the sites with stripe potential $V_0$, whereas the red filled spheres represent the sites without stripe potential $V_0$. The blue filled spheres label the striped rows. Here, we set the period of charge stripes P to 4, based on the observation of the stripe period in cuprates from many experiments \cite{Wu2011,doi:10.1073/pnas.1614247113,PhysRevX.9.021021}.

The Hamiltonian of the extended Hubbard model on the bilayer square lattice is defined as follows
\begin{align}
H = & -t^{\phantom{\dagger}}\sum_{\langle\bf{i,j}\rangle\bf{d}\sigma}
(c^{\dagger}_{\bf{i}\bf{d}\sigma}c^{\phantom{\dagger}}_{\bf{j}\bf{d}\sigma} +c^{\dagger}_{\bf{j}\bf{d}\sigma}c^{\phantom{\dagger}}_{\bf{i}\bf{d}\sigma})
+U\sum_{\bf{i}\bf{d}} n_{\bf{i}\bf{d}\uparrow}n_{\bf{i}\bf{d}\downarrow}\nonumber\\
&-t_{\perp\phantom{\dagger}}\sum_{\bf{i}\sigma}
(c^{\dagger}_{\bf{i}\bf{1}\sigma}c^{\phantom{\dagger}}_{\bf{i}\bf{2}\sigma} +c^{\dagger}_{\bf{i}\bf{2}\sigma}c^{\phantom{\dagger}}_{\bf{i}\bf{1}\sigma})
+V_{0}\sum_{\bf{i}\in stripe}n_{\bf{i}\bf{d}}\nonumber\\
& -t'^{\phantom{\dagger}}\sum_{\langle\langle\bf{i,j}\rangle\rangle\bf{d}\sigma}
(c^{\dagger}_{\bf{i}\bf{d}\sigma}c^{\phantom{\dagger}}_{\bf{j}\bf{d}\sigma}+c^{\dagger}_{\bf{j}\bf{d}\sigma}c^{\phantom{\dagger}}_{\bf{i}\bf{d}\sigma})+V\sum_{\langle\bf{i,j}\rangle\bf{d}} n_{\bf{i}\bf{d}}n_{\bf{j}\bf{d}}
\label{eq:model}
\end{align}

Here, $c^{\dagger}_{\bf{i}\bf{d}\sigma}(c^{\phantom{\dagger}}_{\bf{i}\bf{d}\sigma})$ creates(annihilates) electrons at layer $\bf{d}$ site $\bf{i}$ with spin $\sigma(\sigma=\uparrow,\downarrow)$, and $n_{\bf{i}\bf{d}\sigma}=c^{\dagger}_{\bf{i}\bf{d}\sigma}c^{\phantom{\dagger}}_{\bf{i}\bf{d}\sigma}$ is the occupation number operator for spin $\sigma$. We consider the intralayer hopping $t$ between the nearest neighbor lattice sites $\langle \bf{i,j}\rangle$, the intralayer hopping $t'$ between the next-nearest-neighbor lattice sites $\langle\langle \bf{i,j}\rangle\rangle$ and the interlayer hopping kinetic energy $t_\perp$. $U$ is the onsite Coulomb repulsion interaction, while $V$ represents the interaction between the nearest neighbors. 
Here, we consider the application of a chemical potential of 0 on the interstriped rows while imposing a chemical potential of $V_0$ on the striped rows. The effect of $V_0$ is to facilitate the flow of electrons from the striped rows to the interstriped rows, without altering the overall electron concentration. A sufficiently large $V_0$ is necessary to achieve half-filling of electron concentration in the interstriped rows\cite{PhysRevB.86.184506, Chen2024}. Furthermore, it is important to note that the critical factor influencing electron distribution is not the absolute value of the local chemical potential $V_0$, but the difference in chemical potential between the striped and interstriped rows.
Here, we set $t_\perp=0.2$ and $U=4$. 

We adopt the Constrained Path Quantum Monte Carlo method to study magnetic properties and superconductivity. The CPMC method is a type of projection quantum Monte Carlo method. The ground state wave function is obtained by continuously projecting the initial wave function$|\varphi_{g}\rangle=\lim\limits_{\beta\rightarrow\infty}e^{-\beta\hat{H}}|\varphi_{T}\rangle$, which are used to calculate the Green’s function and thereby measure physical quantities. The operator of the projection wave function is a four operator, and CPMC introduce an auxiliary field to transform it into the desired form. The Monte Carlo process can be regard as a random flip of the auxiliary field, which is also a random walk. The infamous sign problem is caused by random walks. CPMC effectively avoids sign problems by constraining the path approximation, restricting the random walk of the samples in a single degenerate space. CPMC also introduces importance sampling, increasing the sampling probability of the samples with a greater impact on the system to improve iterative efficiency. For more information about CPMC, please see reference \cite{NGUYEN20143344,PhysRevB.55.7464,PhysRevLett.74.3652,PhysRevB.107.245106}.

In cuprate superconductors, the superconductivity arises from $d$-wave pairing. The $d$-wave pairing function is written as
\begin{align}
&C_{d}(\bf{i},\bf{j})=\langle\Delta_{\bf{i}}\Delta^{\dagger}_{ \bf{j}}+\Delta_{\bf{j}}\Delta^{\dagger}_{\bf{i}}\rangle \nonumber\\
&\Delta^{\dagger}_{ \bf{j}}=c^{\dagger}_{\bf{j}\uparrow}(c^{\dagger}_{\bf{j}+\hat{x}\downarrow}-c^{\dagger}_{\bf{j}+\hat{y}\downarrow}
+c^{\dagger}_{\bf{j}-\hat{x}\downarrow}-c^{\dagger}_{\bf{j}-\hat{y}\downarrow})
\label{eq:Saf}
\end{align}
The correlation between electrons leads to superconducting pairing. $C_{d}$ contains the unpairing term, which means that there is no interaction between electrons. This term does not contain important information and can interfere with the expression of superconducting pairing \cite{PhysRevB.84.121410,PhysRevLett.110.107002,Huang2018AntiferromagneticallyOM}. Therefore, we are more concerned with the effective pairing correlations generated from the interactions. The effective pairing function is written as

\begin{equation}
P_{d}(\bf{i},\bf{j}) = C_{d}(\bf{i},\bf{j}) - \widetilde{C}_{d}(\bf{i},\bf{j}),
\label{Vpaire}
\end{equation}
The unpairing term $\widetilde{C}_{d}(\bf{i},\bf{j})$ is an uncorrelated single-particle contribution, which is achieved by replacing $\langle c_{i-\hat{x}\downarrow}c_{i\uparrow}c_{j\uparrow}^{\dagger}c_{j+\hat{x}\downarrow}^{\dagger}\rangle$ with $\langle c_{j+\hat{x}\downarrow}^{\dagger}c_{i-\hat{x}\downarrow}\rangle \langle c_{j\uparrow}^{\dagger}c_{i\uparrow}\rangle$. The positive value of $P_{d}(\bf{i},\bf{j})$ suggests the potential for superconductivity driven by electron-electron correlations. Moreover, a larger magnitude of $P_{d}(\bf{i},\bf{j})$ indicates a stronger $d$-wave pairing interaction.

To study magnetic properties, we define the spin correlation $C_{spin}$ as
\begin{align}
C_{\text{spin}}(\bf{i}) =&
\langle S^{-}_{\bf{j}+\bf{i}}S^{+}_{\bf{j}}\rangle \nonumber\\
S^{+}_{\bf{j}}=&c^{\dagger}_{\bf{j}\uparrow}c^{\phantom{\dagger}}_{\bf{j}\downarrow}
\label{eq:condform}
\end{align}
Here $S^{+}_{\bf{j}}(S^{-}_{\bf{j}})$ is the spin-raising (spin-lowering) operator at $\bf{j}$ site. $\bf{i}$ in the $C_{\text{spin}}(\bf{i})$ refers to the distance between the spin correlated sites. If $C_{\text{spin}}(\bf{i})>0$, the spin direction at the $\bf{j+i}$ site is the same as the spin direction at the $\bf{j}$ site. If $C_{\text{spin}}(\bf{i})<0$, the spin direction at the $\bf{i}$ site is opposite to the spin direction at the $\bf{j+i}$ site. Through $C_{\text{spin}}$, we can determine the magnetic behavior of the system, for example, whether the system exhibits antiferromagnetic order and whether or not the system exhibits SDW. 

\section{Results and discussion}

\begin{figure}[tbp]
\includegraphics[scale=0.4]{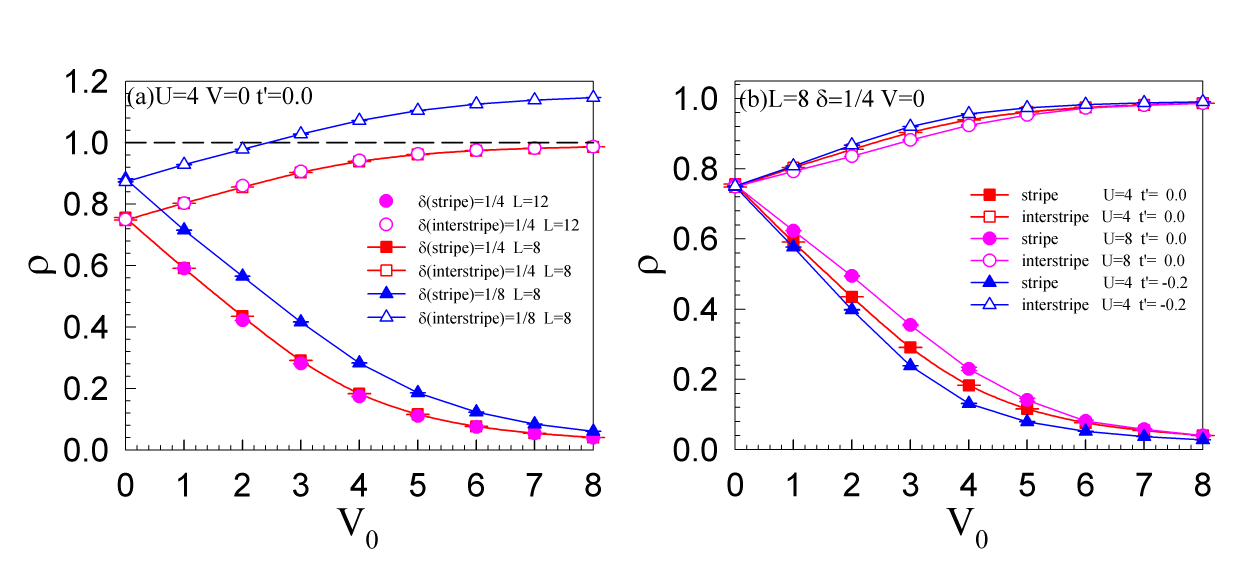}
\caption{Electron densities of both the striped rows and the interstriped rows are shown as a function of $V_0$ (a)for different lattice sizes and hole doping concentrations (b)for different $U$ and $t'$. The empty symbols represent the interstriped rows and the filled symbols represent the striped rows.
}
\label{Fig2}
\end{figure}

In Fig. \ref{Fig2}, we present the results for the electron density of both the striped rows and the interstriped rows. The filled symbols represent the striped rows, while the empty symbols represent the interstriped rows, as illustrated in Fig. \ref{Fig1}. For $V_0=0$, the electron densities of both types of rows are identical. As $V_0$ increases, a decrease in the electron density of the striped rows is accompanied by an increase in the electron density of the interstriped rows. This indicates that $V_0$ induces stripes by transferring electrons from the striped rows to the interstriped rows. Notably, at the hole doping $\delta_h=1/4$, the electron density on the interstriped rows approaches 1 when $V_0=8$, corresponding to a half-filled state. Conversely, at the hole doping $\delta_h=1/8$, half-filling occurs for the interstriped rows when $V_0\approx3$. Furthermore, comparison of the red and pink curves in Fig. \ref{Fig1}(a) clearly shows that results obtained for different lattice sizes are nearly indistinguishable, suggesting that lattice size does not significantly influence the electron density. In addition, we explored the effects of $U$ and $t'$ on the variation of electron concentration in Fig. \ref{Fig2}(b). We found that, although the introduction of larger $U$ and $t'$ slightly influence the system, they do not alter the overall trend of the electron concentration variation.

\begin{figure}[tbp]
\includegraphics[scale=0.4]{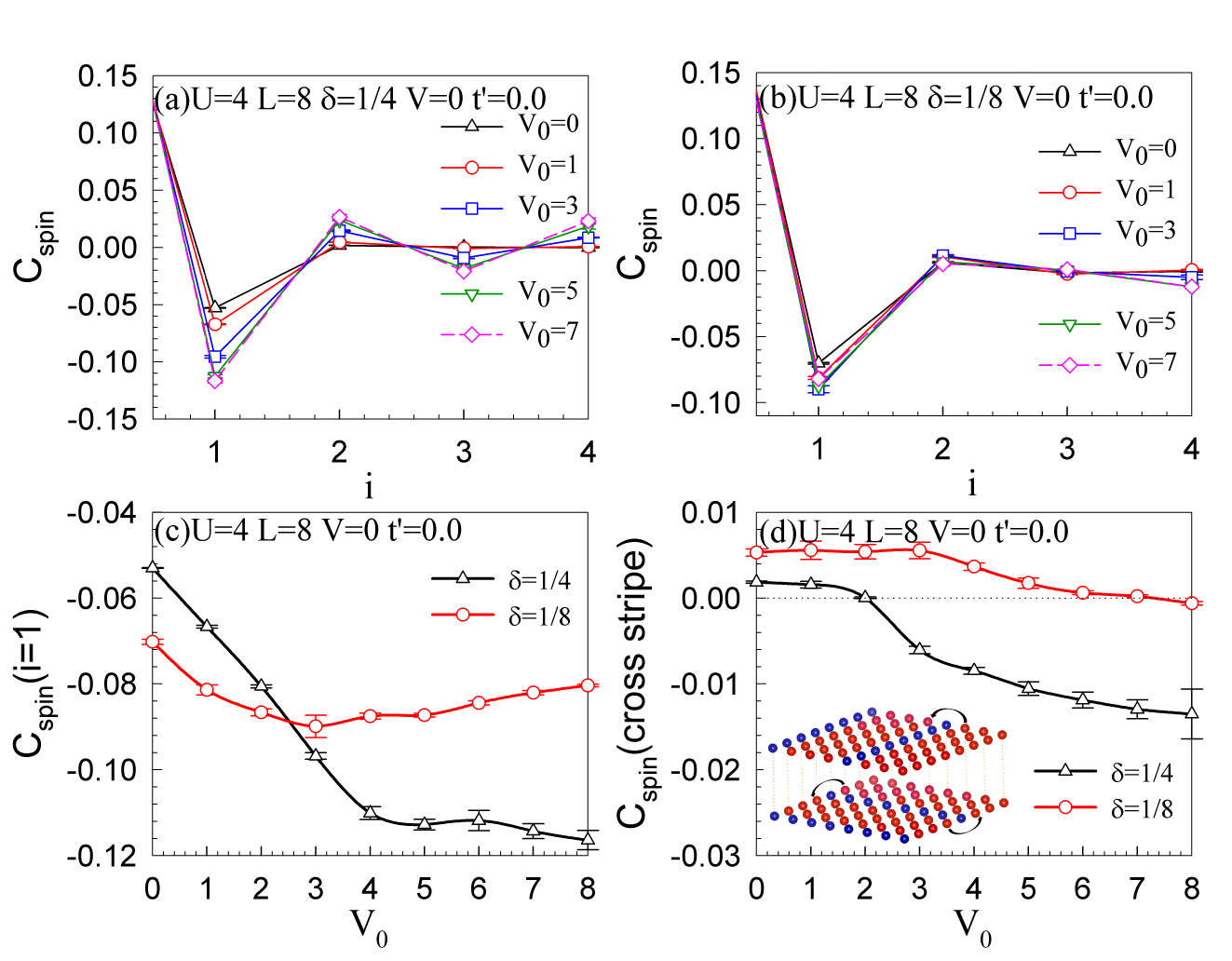}
\caption{ Spin correlation function of the crossed bilayer square model. $C_{\text{spin}}(\text{i})$ at various values of $V_0$ along the interstriped rows (a) at hole doping $\delta_h=1/4$. (b) at hole doping $\delta_h=1/8$. (c) $C_{\text{spin}}(\text{i})$, the spin correlation function for fixed distance i=1 on interstriped rows, as a function of $V_0$ for different hole doping concentrations. (d) $C_{\text{spin}}(\text{cross stripe})$,which represents the spin correlation function between pairs of sites traversing the stripes, as a function of $V_0$ for different hole doping concentrations. Arrows in the inset show the pairs of sites traversing the stripes.
}
\label{Fig3}
\end{figure}

In Fig. \ref{Fig3}, the spin correlation function $C_{\text{spin}}(\text{i})$ for various values of $V_0$ is presented along the interstriped rows at hole doping (a)$\delta_h=1/4$ and (b)$\delta_h=1/8$. It is evident that the spin correlation between nearest sites, $C_{\text{spin}}(\text{i}=1)$, is relatively large, while correlations at other distances approach zero. This indicates that the system exhibits short-range antiferromagnetism for all values of $V_0$ at both hole doping concentrations. We observe that $C_{\text{spin}}(\text{i}=1)$ reaches its maximum when $V_0=7$ for hole doping $\delta_h=1/4$, and when $V_0=3$ for hole doping $\delta_h=1/8$. A possible explanation for this behavior is that the interstriped rows are nearly half-filled in these two scenarios. To further investigate this phenomenon, we present $C_{\text{spin}}(\text{i}=1)$ as a function of $V_0$ in Fig. \ref{Fig3}(c). For hole doping $\delta_h=1/4$, $C_{\text{spin}}(\text{i}=1)$ is monotonically enhanced by increasing $V_0$ as the electron density in the interstriped rows approaches half-filled with rising $V_0$. The maximum value of $C_{\text{spin}}(\text{i}=1)$ occurs at $V_0=8$, which corresponds to a striped potential closest to half-filling. Conversely, at hole doping $\delta_h=1/8$, $C_{\text{spin}}(\text{i}=1)$ increases with small values of $V_0$ but decreases with larger ones. Notably, there exists a critical point precisely at $V_0=3$, corresponding to half-filling. Additionally, we examine the spin correlation along the vertical direction of striped rows. In Fig. \ref{Fig3}(d), we show $C_{\text{spin}}(\text{cross stripe})$, which represents the spin correlation between pairs of sites traversing the stripe (see inset for a sketch), as a function of $V_0$. At both $\delta_h = 1 /4$ and $\delta_h = 1 /8$ hole dopings, $C_{\text{spin}}(\text{cross stripe})$ remains positive when $V_0=0$. However, as $V_0$ increases, $C_{\text{spin}}(\text{cross stripe})$ decreases to a negative value. The sign changes from positive to negative in $C_{\text{spin}}(\text{cross stripe})$, induced by $V_0$, which corresponds to the $\pi$-phase shift of the spins observed experimentally \cite{Tranquada1995EvidenceFS,PhysRevLett.78.338}.

\begin{figure}[tbp]
\includegraphics[scale=0.41]{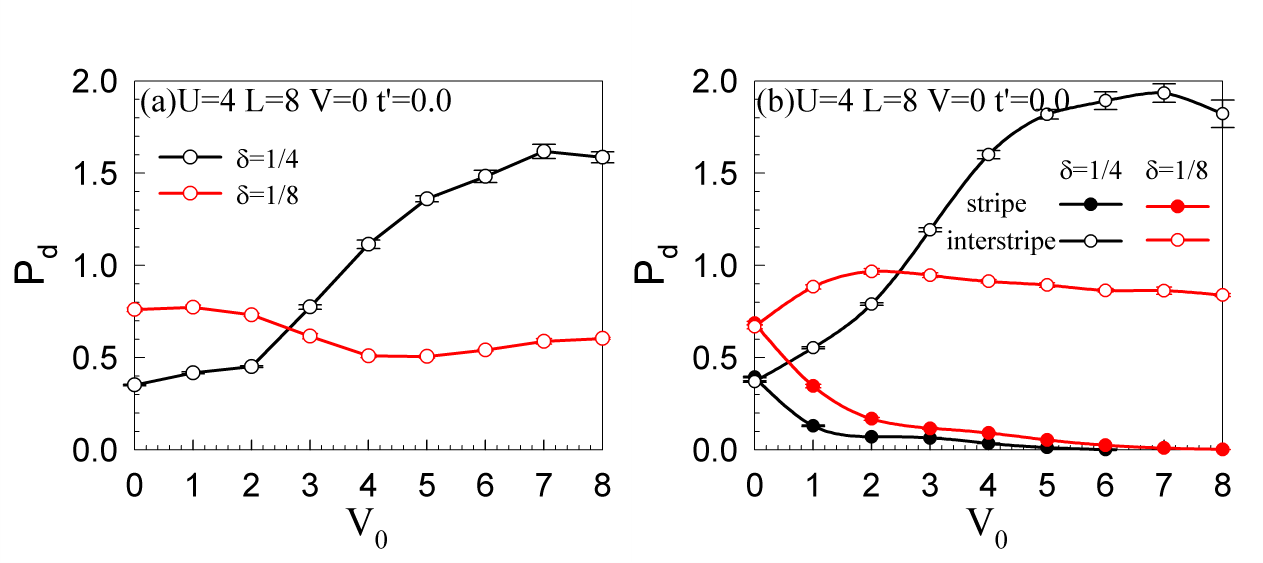}
\caption{ (a) Effective $d$-wave pairing interaction $P_d$ of the entire lattice as a function of $V_0$ at different hole doping concentrations. (b) Effective $d$-wave pairing interaction $P_d$ of both the striped rows and the interstriped rows as a function of $V_0$ at different hole doping concentrations.
}
\label{Fig4}
\end{figure}

Next, we focus on the $d$-wave pairing in the crossed bilayer square lattice. Fig. \ref{Fig4}(a) shows the effective $d$-wave pairing correlation $P_d$ of the entire lattice as a function of $V_0$ at various hole doping concentrations. As $V_0$ increases, $P_d$ is enhanced at hole doping $\delta_h = 1 /4$, whereas $P_d$ is suppressed at hole doping $\delta_h = 1/8$, indicating that hole doping concentration plays a crucial role in the relationship between stripe and $d$-wave pairing.
In Fig. \ref{Fig4}(b), we present the effective $d$-wave pairing correlation $P_d$ of both the striped rows and interstriped rows to provide more details about the impact of the striped potential. When $V_0=0$, the values of $P_d$ are nearly identical for both the striped rows and interstriped rows. As $V_0$ increases, at hole doping $\delta_h = 1 /4$, the effective $d$-wave pairing correlation $P_d$ of the striped rows decreases, while that of the interstriped rows increases. The trend of the effective $d$-wave pairing $P_d$ of interstriped rows and $C_{\text{spin}}(\text{i}=1)$(see in Fig. \ref{Fig3}(c)) both as functions of $V_0$ are perfectly aligned, indicating that the $d$-wave pairing in the system is driven by antiferromagnetism. Furthermore, since the decrease in $P_d$ of the striped rows is not as significant as the increase of the interstriped rows, the d-wave pairing of the entire lattice increases. For hole doping $\delta_h=1/8$, the $d$-wave pairing correlation $P_d$ of the striped rows exhibits behavior consistent with that at $\delta_h=1/4$. However, the $P_d$ of the interstriped rows shows a trend of first increasing and then decreasing, which is also identical to the trend of $C_{\text{spin}}(\text{i}=1)$. At hole doping $\delta_h=1/8$, the decrease in $P_d$ of the striped rows surpasses the initial increase of the interstriped rows, leading to the striped potential $V_0$ suppressing the $d$-wave pairing overall.

\begin{figure}[tbp]
\includegraphics[scale=0.41]{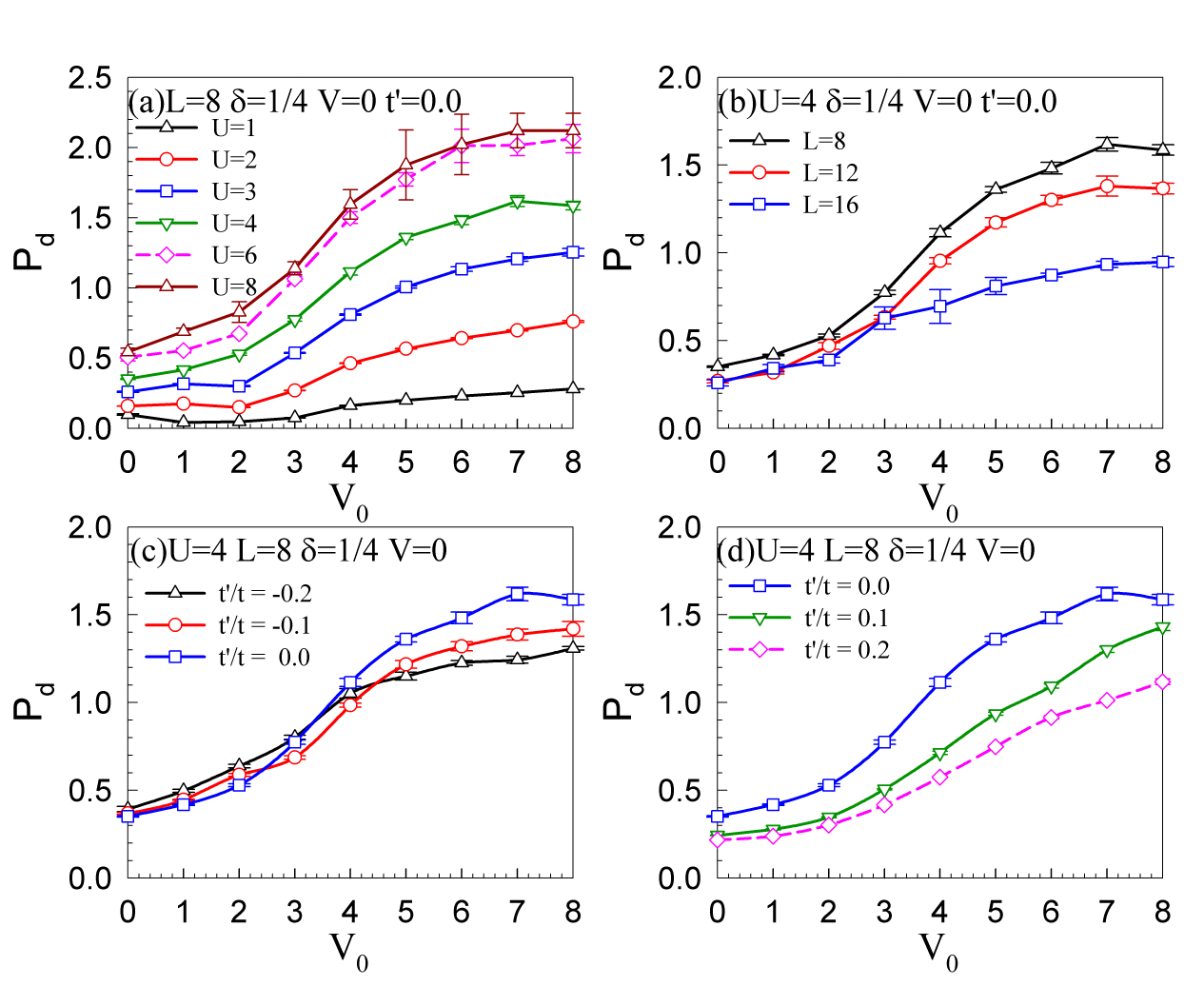}
\caption{ Effective $d$-wave pairing interaction $P_d$ of the entire lattice as a function of $V_0$ (a) at different values of $U$ (b) on different lattice sizes (c) with negative $t'$ (d) with positive $t'$  for hole doping $\delta_h=1/4$
}
\label{Fig5}
\end{figure}

The above results are based on the parameter settings of $U=4$, $L=8$ and $t'= 0.0$. Next, we examine the role of the on-site Coulomb interaction $U$, lattice size $L$ and next-nearest-neighbor hopping $t'$ on $d$-wave pairing. Fig. \ref{Fig5} shows the effective $d$-wave pairing interaction $P_d$ of the entire lattice as a function of $V_0$ (a) at different values of $U$ (b) on different lattice sizes (c) with negative $t'$ (d) with positive $t'$ for hole doping $\delta_h=1/4$. It is observed from Fig. \ref{Fig5}(a) that $P_d$ is enhanced by $V_0$ across all values of the on-site Coulomb interaction $U$. Furthermore, $d$-wave pairing is enhanced with increasing $U$, as the stronger repulsion leads to significant electron-electron correlation. Fig. \ref{Fig5}(b) presents the results of the calculations  conducted on larger lattice sizes ($L=12$ and $L=16$) and shows that the properties of the system are consistent with those obtained at $L=8$. This indicates that results based on a lattice size of $L=8$ are reliable. We further explored the impact of $t'$ on $d$-wave pairing and presented our findings in Fig. \ref{Fig5}(c) and (d). These figures consistently demonstrate that at $\delta = 1/4$, introducing $t'$ does not modify the enhancing effect that stripes have on $d$-wave pairing. Fig. \ref{Fig5}(c) also revealed a more complex relationship when next-nearest-neighbor hopping is negative: in relatively uniform systems or those under small $V_0$, $t'$ slightly enhances $d$-wave pairing; however, in systems under relatively large $V_0$, $t'$ suppresses $d$-wave pairing. However, when next-nearest-neighbor hopping is positive in Fig. \ref{Fig5}(d), $t'$ consistently exhibits a suppressive effect on $d$-wave pairing.

\begin{figure}[tbp]
\includegraphics[scale=0.41]{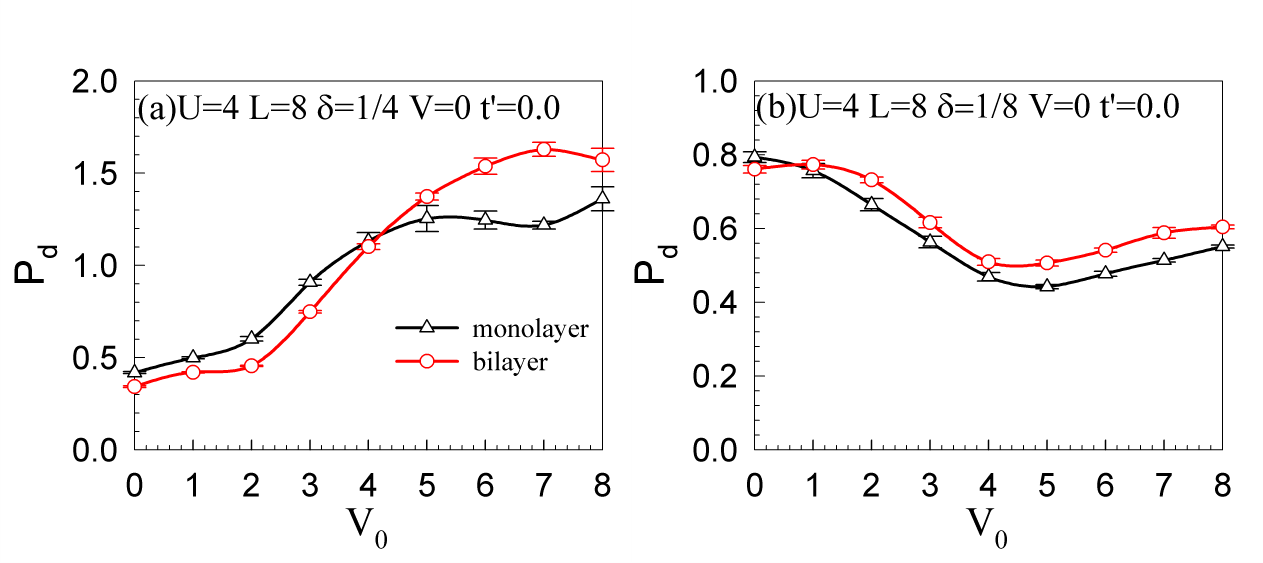}
\caption{ A comparison of the effective $d$-wave pairing interaction $P_d$ between the monolayer and bilayer striped square lattice models at (a) hole doping $\delta_h=1/4$ and (b) hole doping $\delta_h=1/8$.
}
\label{Fig6}
\end{figure}

Next, we analyze the differences between these two models from the perspective of stripes. In Fig. \ref{Fig6}, we compare the effective $d$-wave pairing interaction in monolayer model with that in the bilayer model at hole doping (a)$\delta_h=1/4$ and (b) $\delta_h=1/8$. At both hole doping concentrations, $P_d$ in monolayer striped model is larger than that in the bilayer striped model when the system is homogeneous ($V_0=0$), which is consistent with previous studies \cite{TEWARI20211353895,Gall2021}. However, as $V_0$ increases, the $P_d$ of the bilayer model exceeds that of the monolayer model at both hole doping concentrations. 

\begin{figure}[tbp]
\includegraphics[scale=0.41]{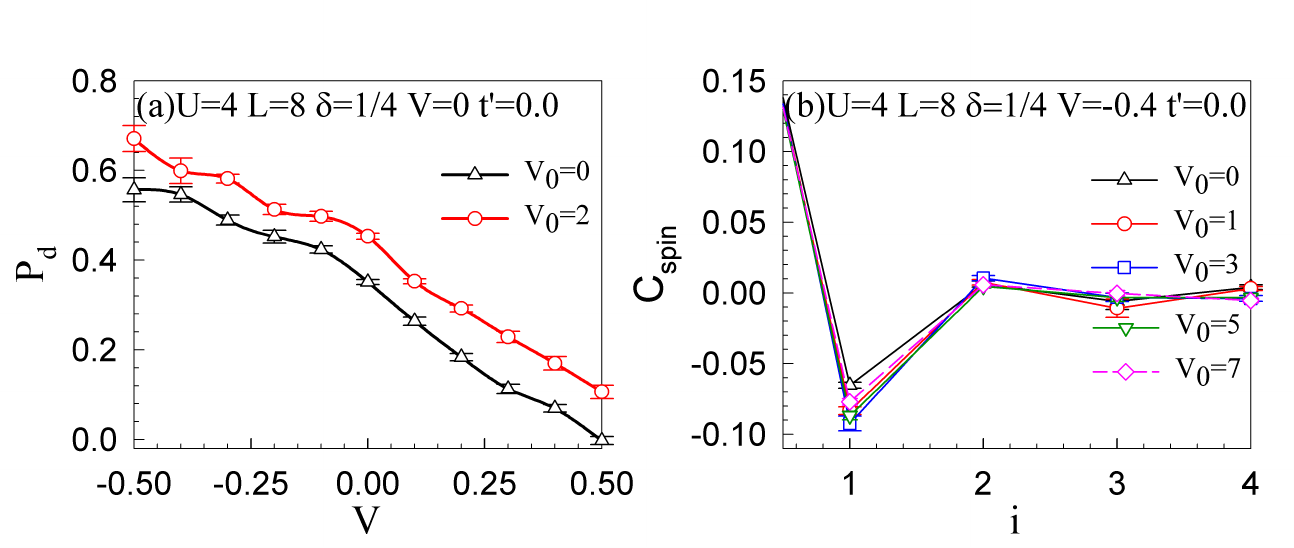}
\caption{ (a)Effective $d$-wave pairing interaction $P_d$ of the entire lattice as a function of the nearest neighbor interaction $V$ at different values of $V_0$. The hole doping concentration of the lattice is fixed at $\delta=1/4$. (b)Spin correlation function of the crossed bilayer square model. $C_{\text{spin}}(\text{i})$ along the interstriped rows at hole doping $\delta_h=1/4$ when $V=-0.4$.  
}
\label{Fig7}
\end{figure}

Recent experimental results of angle-resolved photoemission spectroscopy suggested that an additional nearest neighbor attraction in the Hubbard model may be important for the doped cuprates \cite{doi:10.1126/science.abf5174}. 
Moreover, ab initio calculations indicate that nearest-neighbor repulsive interactions exist in cuprates\cite{PhysRevB.98.134501}, and the presence of such interactions facilitates the formation of charge density waves (CDW)\cite{Abram_2017}.
In Fig. \ref{Fig7}(a), we consider the effect of the nearest neighbor interaction $V$ on the pairing interaction. 
As demonstrated in Fig. \ref{Fig7}(a), the $d$-wave pairing strength is enhanced by negative $V$, while it is diminished by positive $V$. This observation suggests that additional nearest neighbor attraction promotes $d$-wave pairing, whereas additional nearest neighbor repulsion impedes it.
Furthermore, we examine the influence of $V$ on magnetism in Fig. \ref{Fig7}(b), which yields results consistent with those previously discussed.

\section{Conclusions}
In summary, we employ the CPMC method to study the spin correlations and $d$-wave pairing of the crossed bilayer square lattice at striped period $P=4$. We find that the system exhibits short-range antiferromagnetic order, and this antiferromagnetism is strongest when the electron density of the interstriped rows reaches half-filling. Examination of the spin correlation function traversing the striped rows reveals that the spin correlation function undergoes a sign change from positive to negative as $V_0$ increases, indicating a $\pi$-phase shift. As for the $d$-wave pairing, we discover that the striped potential $V_0$ can have both positive and negative effects on the effective $d$-wave pairing. As $V_0$ increases, $P_d$ is enhanced at hole doping $\delta_h = 1 /4$, whereas $P_d$ is suppressed at hole doping $\delta_h = 1/8$. To provide further insight into the impact of the striped potential, we present the effective $d$-wave pairing $P_d$ of both the striped rows and interstriped rows and find that the different effect of striped potential can be explained based on the magnetism of the interstriped rows. Furthermore, we analyze the differences between the monolayer model and the bilayer model from the perspective of stripes. When stripes are introduced on the square lattice, the effective $d$-wave pairing is larger in bilayers compared to monolayers. In the presence of a nearest neighbor attractive interaction $V$, the $d$-wave pairing is promoted.

Some intriguing questions remain open. In the study of cuprates, the momentum structure of the superconducting order parameter and the existence of pair density wave (PDW) states represent central issues\cite{PhysRevLett.99.127003,PhysRevLett.88.117001,PhysRevB.76.140505}. Particularly noteworthy is the $\pi$-phase shift in spin correlations and the stripe phase discussed in this paper, which may facilitate the formation of a $d$-wave pairing with a non-zero wave vector. Based on this, we intend to further investigate the momentum structure of the $d$-wave pairing in future work to achieve a more comprehensive and profound understanding.

\section*{Acknowledgments}
This work was supported by NSFC (12474218) and Beijing Natural Science
Foundation (No. 1242022 and 1252022). The numerical simulations in this work were performed at the HSCC of Beijing Normal University.

\bibliography{reference}

\end{document}